\documentclass[
 aps, pra,
 amsmath,amssymb,
 10pt,
 final,
tightenlines,
 twoside,
 twocolumn,
 floatfix,
 nobibnotes,
nofootinbib,
 superscriptaddress,
showkeys,
 ]
{revtex4-2}

\usepackage[utf8x]{inputenc}
\usepackage[english]{babel}
\usepackage{graphicx}
\usepackage{dcolumn}
\usepackage{bm}
\usepackage{natbib}
\usepackage{float}
\usepackage{hyperref}
\usepackage{slashed}
\usepackage{multirow}
\usepackage{lineno}
\usepackage{booktabs}
\usepackage{silence}
\usepackage{scalerel}

\begin{document}

\selectlanguage{english}



\title{Chromomagnetic dipole moments of light quarks in the Bestest Little Higgs Model}

\author{\firstname{T.}~\surname{Cisneros-Pérez} }
 \email{tzihue@gmail.com}
 \affiliation{Unidad Académica de Ciencias Químicas, Universidad Autónoma de  Zacatecas,Apartado Postal C-585, 98060 Zacatecas, Mexico. }

\author{\firstname{E.}~\surname{Cruz-Albaro}}
 \email{elicruzalbaro88@gmail.com}
 \affiliation{Facultad de Física, Universidad Autónoma de Zacatecas, Apartado Postal C-580, 98060 Zacatecas, Mexico}

\author{\firstname{A. Y.}~\surname{Ojeda-Castañeda}}
 \email{angelica.ojeda@fisica.uaz.edu.mx}
 \affiliation{Facultad de Física, Universidad Autónoma de Zacatecas, Apartado Postal C-580, 98060 Zacatecas, Mexico}

\author{\firstname{S. E.}~\surname{Solís-Núñez}}
 \email{sara.solis@fisica.uaz.edu.mx}
 \affiliation{Facultad de Física, Universidad Autónoma de Zacatecas, Apartado Postal C-580, 98060 Zacatecas, Mexico}

\begin{abstract}
In this paper, we research into the anomalous Chromomagnetic Dipole Moment (CMDM), denoted as $\hat{\mu}_{q}^{BLHM}$, of the light quarks $q=(u, c, d, s, b)$ within the framework of the Bestest Little Higgs Model (BLHM) as an extension of the Standard Model (SM). Our investigation encompasses novel interactions among the light quarks, the heavy quark $B$, and the heavy bosons $(W^{\prime\pm}, H^{\pm}, \phi^{\pm}, \eta^{\pm})$, incorporating the extended Cabibbo-Kobayashi-Maskawa (CKM) matrix characteristic of the BLHM. We thoroughly explore the permissible parameter space, yielding a spectrum of CMDM values ranging from $10^{-10}$ to $10^{-3}$.
\end{abstract}

\maketitle

\section{INTRODUCTION}

The CMDM of the top quark has been extensively calculated theoretically within the framework of the SM in several studies \cite{Martinez:1996cy,Martinez:2001qs,Martinez:2007qf,BuarqueFranzosi:2015jrv,Aranda:2020tox}, with the most accurate experimental measurement reported in \cite{ATLAS:2021upq}. Conversely, extended models in the literature have explored both the CMDM and the chromoelectric dipole moment (CEDM) \cite{Ding:2008nh,Hernandez-Juarez:2018uow,Hernandez-Juarez:2020gxp,Aranda:2021kza}, yielding results and theoretical implications that vary depending on each model beyond the SM (BSM). The prevalence of studies on the CMDM and CEDM of the top quark over the light quarks of the SM in BSM is precisely due to the magnitude of its mass and the experimental stage of the last decade in which interactions with heavy particles above 1 TeV were expected to be found.
Several experimental reports show the development towards increasingly higher energies and therefore towards hypothetical particles that even exceed 5 TeV \cite{rappoccio2019experimental}.
In this regard, calculating the CMDMs for the light quarks $(u,c,d,s,b)$ within the framework of certain BSM scenarios may seem unnecessary in the absence of experimental measurements. However, it is noteworthy that a central value for $\hat{\mu}_t^{SM}$ already exists, and the CEDM is bounded. This contributes uniquely to the understanding within the BLHM, complementing existing studies on the subject 
\cite{Aranda:2021kza,Cruz-Albaro:2022kty,Cruz-Albaro:2022lks,Cruz-Albaro:2023pah}. The CMDM of the light quarks was calculated within the framework of the SM in \cite{Choudhury2015} for the spacelike value $(q^2=-m_Z^2)$, while in \cite{moyotl2021} it is studied for the timelike value $(q^2=+m^2_Z)$. In the article \cite{Montano-Dominguez:2021eeg}, the authors recalculate them for both values $(q^2=\pm m^2_Z)$, providing detailed results for the individual contributions received by each light quark from the chromodynamic and electroweak parts. In this study, we calculate the CMDM of the light quarks $(u,c,d,s,b)$ in the BLHM (in \cite{Cisneros-Perez2023} has been shown that the CEDM does not exist in this model) for both values $(q^2=\pm m^2_Z)$ of the off-shell gluon and with the on-shell quarks. In this case, we find differences in both scenarios, mainly for values greater than 2 TeV in the symmetry breaking scale of the model. 
The BLHM is a BSM model that has not been explored as much as other models in the Little Higgs Model (LHM) family, as it considers larger masses whose experimental observation was more difficult. However, in the latest CERN update, we could expect results from models like the BLHM to stand out. Among other objectives, the BLHM was constructed \cite{schmaltz2010bestest} to address certain issues in the LHM family such as divergent singlets, masses of heavy bosons smaller than masses of heavy quarks, and custodial symmetry, among others \cite{Schmaltz:2008vd}. A unique aspect of this model is its modular structure, which requires two distinct breaking scales, $f$ and $F$, with the condition $F>f$. In this way, the heavy quarks depend on the scale $f$, and the heavy gauge bosons depend on both $f$ and $F$, where $F$ can be as large as necessary. The BLHM also offers a highly enriched phenomenology due to its fermionic and bosonic content, whose contributions to the light quarks may provide interesting sightings for signals of new physics in the leading medium-term planned experiments.

The outline of this paper is as follows. In Section \ref{seccion2}, we provide a brief introduction to the BLHM. In Section \ref{seccion3}, we discuss the effective Lagrangian containing the magnetic dipole moment form factor. In Section \ref{seccion4}, we describe the parameter space of the BLHM and the experimental limits that constrain it. In Section \ref{seccion5}, we develop the phenomenology of the CMDM of light quarks and present our results. Finally, in Section \ref{seccion6}, we provide our conclusions. In Appendix \ref{feynrules} we show the new Feynman rules used.

\section{A brief review of the BLHM}
\label{seccion2}

The BLHM~\cite{schmaltz2010bestest} comes from the group $SO(6)_A \times SO(6)_B$, which experiences a breaking at the scale $f$ towards $SO(6)_V$ when the non-linear sigma field $\Sigma$ acquires a vacuum expectation value (VEV), $\langle\Sigma\rangle = 1$,

\begin{equation}\label{Sigma}
\Sigma=e^{i\Pi/f}e^{2i\Pi_h/f}e^{i\Pi/f}.
\end{equation}

\noindent This results in 15 pseudo-Nambu Goldstone bosons, parameterized by means of the electroweak triplet $\phi^a$ with zero hypercharges $(a=1,2,3)$ and the triplet $\eta^a$, where $(\eta_1,\eta_2)$ form a complex singlet with hypercharge, and $\eta_3$ becomes a real singlet. The scalar field $\sigma$ is required to generate a collective quartic coupling~\cite{schmaltz2010bestest} and $T_{L,R}^a$ denote the generators of $SU(2)_L$ and $SU(2)_R$ groups, Eq. \ref{triples},

\begin{equation}\label{triples}
\Pi=\begin{pmatrix}
\phi_a T^a_L+\eta_aT^a_R & 0 & 0\\
0 & 0 & i\sigma/\sqrt{2}\\
0 & -i\sigma/\sqrt{2} & 0
\end{pmatrix}.
\end{equation}

\noindent In matrix $\Pi_h$, the  $h_i^T=(h_{i1},h_{i2},h_{i3},h_{i4})$, $(i=1,2)$, represent Higgs quadruplets of $SO(4)$,

\begin{equation}
\Pi_h=\begin{pmatrix}
0_{4\times4} & h_1 & h_2\\
-h^T_1 & 0 & 0\\
-h^T_2 & 0 & 0
\end{pmatrix}.
\end{equation}

%
%

\subsection{Scalar sector}

In the context of the BLHM, two operators are necessary to induce the quartic coupling of the Higgs via collective symmetry breaking; neither of these operators alone enables the Higgs to develop a potential:

\begin{eqnarray}
P_5=diag(0,0,0,0,1,0),\\\nonumber
 P_6=diag(0,0,0,0,0,1).
\end{eqnarray}

\noindent Thus, we can express the quartic potential as~\cite{schmaltz2010bestest}

\begin{eqnarray}\label{potVq}\nonumber
 V_q&=&\frac{1}{4}\lambda_{65}f^4Tr(P_6\Sigma P_5\Sigma^T)\\\nonumber
 &+&\frac{1}{4}\lambda_{56}f^4Tr(P_5\Sigma P_6 \Sigma^T)\\
 &=&\frac{1}{4}\lambda_{65}f^4(\Sigma_{65})^2+\frac{1}{4}\lambda_{56}f^4(\Sigma_{56})^2,
\end{eqnarray}

\noindent where $\lambda_{56}$ and $\lambda_{65}$ are non-zero coefficients necessary to realize collective symmetry breaking and produce a quartic coupling for the Higgs.

The initial segment of Eq. (\ref{potVq}) induces a breaking of $SO(6)_A \times SO(6)_B$ to $SO(5)_{A5} \times SO(5)_{B6}$, whereby $SO(5)_{A5}$ prohibits $h_1$ from acquiring a potential, and $SO(5)_{B6}$ performs the same function for $h_2$. The latter portion of Eq. (\ref{potVq}) leads to a breaking of $SO(6)_A \times SO(6)_B$ to $SO(5)_{A6} \times SO(5)_{B5}$.
If we expand Eq. (\ref{Sigma}) in powers of $1/f$ and substitute it into Eq. (\ref{potVq}), we obtain

\begin{eqnarray}\label{potVqSerie}
 V_q=\frac{\lambda_{65}}{2}\left(f\sigma-\frac{1}{\sqrt{2}}h_1^Th_2+\dots\right)^2\\\nonumber
 +\frac{\lambda_{56}}{2}\left(f\sigma+\frac{1}{\sqrt{2}}h_1^Th_2+\dots\right)^2.
\end{eqnarray}

\noindent This potential generates a mass for $\sigma$,
\begin{equation}\label{masa-sigma}
    m_{\sigma}^2=(\lambda_{65}+\lambda_{56})f^2.
\end{equation}
From Eq.~(\ref{potVqSerie}), it appears that each term individually contributes to the generation of a quartic coupling for the Higgs fields. However, this effect can be nullified by redefining the field $\sigma$ as $\pm \frac{h^{T}_1 h_2}{\sqrt{2} f}$, where the positive and negative signs correspond to the first and second operators in Eq.~(\ref{potVqSerie}), respectively.
In combination, however, the two expressions in Eq.~(\ref{potVqSerie}) yield a quartic Higgs potential at tree level; this occurs subsequent to integrating out $\sigma$~\cite{schmaltz2010bestest,Schmaltz:2008vd,phdthesis}:

\begin{equation}
 V_q=\frac{\lambda_{56} \lambda_{65}}{\lambda_{56}+ \lambda_{65}} \left(h_1^Th_2 \right)^2= \frac{1}{2}\lambda_{0}\left(h_1^Th_2 \right)^2.
\end{equation}

\noindent The expression obtained has the desired form of a collective quartic potential~\cite{schmaltz2010bestest,Schmaltz:2008vd}.
Thus, we derive a quartic collective potential form that is dependent on two distinct couplings~\cite{schmaltz2010bestest}. It can be noted that $\lambda_0$ will be zero whenever $\lambda_{56}$, $\lambda_{65}$, or both are zero. This exemplifies the concept of collective symmetry breaking.

Excluding gauge interactions, not all scalars acquire mass. Consequently, it becomes necessary to introduce the potential,

\begin{eqnarray}\label{potVs}\nonumber
 V_s&=&-\frac{f^2}{4}m_4^2Tr\left(\Delta^{\dagger}M_{26}\Sigma M^{\dagger}_{26}+\Delta M_{26}\Sigma^{\dagger}M^{\dagger}_{26}\right)\\
 &&-\frac{f^2}{4}\left(m_5^2\Sigma_{55}+m_6^2\Sigma_{66}\right),
\end{eqnarray}

\noindent where $m_4$, $m_5$, and $m_6$ represent mass parameters, and $(\Sigma_{55},\Sigma_{66})$ denote the matrix elements from Eq. (\ref{Sigma}). In this context, $M_{26}$ is a matrix that contracts the $SU(2)$ indices of $\Delta$ with the $SO(6)$ indices of $\Sigma$,

\begin{equation}
M_{26}=\frac{1}{\sqrt{2}}
\begin{pmatrix}
 0 & 0 & 1 & i & 0 & 0\\
 1 & -i & 0 & 0 & 0 & 0
\end{pmatrix}.
\end{equation}

\noindent The operator $\Delta$ originates from a global symmetry $SU(2)_C \times SU(2)_D$, which is spontaneously broken to a diagonal $SU(2)$ at the scale $F>f$ upon acquiring a VEV, $\langle\Delta\rangle = 1$. We can parameterize it in the form
\begin{equation}
 \Delta=e^{2i\Pi_d/F},\hspace{0.3cm}\Pi_d=\chi_a\frac{\tau_a}{2}\hspace{0.3cm}(a=1,2,3),
\end{equation}
where the matrix $\Pi_d$ incorporates the scalars of the triplet $\chi_a$ which undergo mixing with the triplet $\phi_a$. $\tau_a$ denotes the Pauli matrices. $\Delta$ is linked to $\Sigma$ in a manner such that the diagonal subgroup of $SU(2)_A \times SU(2)_B \subset SO(6)_A \times SO(6)_B$ is recognized as the SM $SU(2)_L$ group.
If we expand the operator $\Delta$ in powers of $1/F$ and substitute it into Eq. (\ref{potVs}), we obtain
\begin{equation}
 V_s=\frac{1}{2}\left(m^2_{\phi}\phi^2_a+m^2_{\eta}\eta^2_a+m^2_1h^T_1h_1+m^2_2h^T_2h_2\right),
\end{equation}
where
\begin{eqnarray}
 m^2_{\phi}&=&m^2_{\eta}=m^2_4,\\\nonumber
 m^2_1&=&\frac{1}{2}(m^2_4+m^2_5),\\\nonumber
 m^2_2&=&\frac{1}{2}(m^2_4+m^2_6).
\end{eqnarray}
To trigger EWSB, the next potential term is introduced\cite{schmaltz2010bestest}:
\begin{equation}
 V_{B_{\mu}}=m_{56}^2f^2\Sigma_{56}+m_{65}^2f^2\Sigma_{65},
\end{equation}
where $m_{56}$ and $m_{65}$ are masses terms that correspond to the matrix elements $\Sigma_{56}$ and $\Sigma_{65}$, respectively.
Finally, we have the complete scalar potential,
\begin{equation}\label{pEscalar}
 V=V_q+V_s+V_{B_{\mu}}.
\end{equation}

\noindent A potential for the Higgs doublets is necessary; hence, we minimize Eq. (\ref{pEscalar}) with respect to $\sigma$ and then substitute the resulting expression back into Eq. (\ref{pEscalar}), yielding the following expression:

\begin{eqnarray}\label{potVH}\nonumber
 V_H&=&\frac{1}{2}\Big[m_1^2h_1^Th_1+m_2^2h_2^Th_2-2B_{\mu}h_1^Th_2\\
 &+&\lambda_0(h_1^Th_2)^2\Big],
\end{eqnarray}
where

\begin{equation}\label{potB}
 B_{\mu}=2\frac{\lambda_{56}m_{65}^2+\lambda_{65}m_{56}^2}{\lambda_{56}+\lambda_{65}}.
\end{equation}
\noindent The potential (\ref{potVH}) attains a minimum when $m_1m_2 > 0$, and EWSB necessitates that $B_{\mu} > m_1m_2$. It is noteworthy that the term $B_{\mu}$ vanishes if either $\lambda_{56}=0$, $\lambda_{65}=0$, or both are zero in Eq. (\ref{potB}). Following EWSB, the Higgs doublets acquire VEVs given by

\begin{equation}\label{aches}
 \langle h_1\rangle=v_1,\hspace{0.3cm}\langle h_2\rangle=v_2.
\end{equation}

\noindent The two terms in (\ref{aches}) are required to minimize Eq. (\ref{potVH}), leading to the subsequent relationships:

\begin{eqnarray}
 v_1^2=\frac{1}{\lambda_0}\frac{m_2}{m_1}(B_{\mu}-m_1m_2),\\
 v_2^2=\frac{1}{\lambda_0}\frac{m_1}{m_2}(B_{\mu}-m_1m_2),
\end{eqnarray}
and it is defined the $\beta$ angle between $v_1$ and $v_2$~\cite{schmaltz2010bestest}, such that,
\begin{equation}
\tan\beta=\frac{\langle h_{11}\rangle}{\langle h_{21}\rangle}=\frac{v_1}{v_2}=\frac{m_2}{m_1},
\end{equation}
in this way, we have
\begin{eqnarray}
v^2&=&v_1^2+v_2^2\\\nonumber
&=&\frac{1}{\lambda_0}\left(\frac{m_1^2+m_2^2}{m_1m_2}\right)(B_{\mu}-m_1m_2)\\\nonumber
&\simeq&(246\;GeV)^2.
\end{eqnarray}
Following EWSB, the scalar sector~\citep{schmaltz2010bestest,phdthesis} gives rise to massive states such as $h^0$ (the SM Higgs), $A^0$, $H^{\pm}$, and $H^0$, each with their respective masses:

\begin{eqnarray}
\label{masa-esc1}
&&m^2_{G^0}=m^2_{G^{\pm}}=0,\\
\label{masa-esc2}
&&m^2_{A^0}=m^2_{H^{\pm}}=m^2_1+m^2_2,\label{masa-A0}\\
\label{masa-esc3}
&&m^2_{h^0,H^0}=\frac{B_{\mu}}{\sin2\beta}\label{masa-H0}\\\nonumber
\mp&&\sqrt{\frac{B^2_{\mu}}{\sin^22\beta}-2\lambda_0\beta_{\mu} v^2\sin2\beta+\lambda_0^2v^4\sin^22\beta},
\end{eqnarray}
where $G^0$ and $G^{\pm}$ represent the Goldstone bosons that are absorbed to confer masses to the $W^{\pm}$ and $Z$ bosons of the SM.


\subsection{Gauge boson sector}
The Lagrangian including the gauge kinetic terms is provided by,~\citep{schmaltz2010bestest,phdthesis},

\begin{equation}\label{lag-norma}
\mathcal{L}=\frac{f^2}{8}Tr\left(D_{\mu}\Sigma^{\dag}D^{\mu}\Sigma\right)+\frac{F^2}{4}Tr\left(D_{\mu}\Delta^{\dag}D^{\mu}\Delta\right),
\end{equation}
where $D_{\mu}\Sigma$ and $D_{\mu}\Delta$ are covariant derivatives,

\begin{eqnarray}
 D_{\mu}\Sigma&=&i\sum_a\left(g_AA_{1\mu}^aT_L^a\Sigma-g_BA_{2\mu}^a\Sigma T^a_L\right)\\\nonumber
 &+&ig'B_3\left(T^3_R\Sigma-\Sigma T^3_R\right),\\
 D_{\mu}\Delta&=&\frac{i}{2}\sum_a\left(g_AA^a_{1\mu}\tau_a\Delta-g_BA^a_{2\mu}\Delta\tau_a\right),
 \end{eqnarray}

\noindent where $(A^a_{1\mu},A^a_{2\mu})$ represent eigenstates of the gauge bosons, $g'$ is the coupling constant of $U(1)_Y$, and $g$ is the coupling constant of $SU(2)_L$. These are related to the couplings $g_A$ and $g_B$ of $SU(2)_A \times SU(2)_B$ as follows:

\begin{eqnarray}\label{acoplesSU}
g=\frac{g_Ag_B}{\sqrt{g_A^2+g_B^2}},\\
s_g=\sin\theta_g=\frac{g_A}{\sqrt{g_A^2+g_B^2}},\\
c_g=\cos\theta_g=\frac{g_B}{\sqrt{g_A^2+g_B^2}},
\end{eqnarray}
here, $\theta_g$ is the mixing angle, and if $g_A = g_B$, then $\tan\theta_g = 1$.

In the context of the BLHM, the masses of both the heavy gauge bosons $W^{\prime\pm}$, $Z'$, and those of the SM bosons are also generated~\citep{schmaltz2010bestest,phdthesis},

\begin{eqnarray}
\label{masa-zp}
m_{Z'}^2&=&\frac{1}{4}(g_A^2+g_B^2)(f^2+F^2)-\frac{1}{4}g^2v^2\\\nonumber
&+&\Bigg(2g^2+\frac{3f^2}{f^2+F^2}\\\nonumber
&\times&(g^2+g'\,^2)(s_g^2-c_g^2)\Bigg)\frac{v^4}{48f^2},\\
\label{masa-wp}
m_{W'}^2&=&\frac{1}{4}(g_A^2+g_B^2)(f^2+F^2)-m_W^2.
\end{eqnarray}


\subsection{Fermion sector}

The Lagrangian that governs the fermion sector of the BLHM is provided by,~\cite{schmaltz2010bestest},

\begin{eqnarray}
\label{lag-yuk}
\mathcal{L}_t&=&y_1fQ^TS\,\Sigma\, SU^c+y_2fQ_a^{\prime T}\Sigma\,U^c\\\nonumber
&+&y_3fQ^T\Sigma\, U_5^{\prime c}+y_bfq_3^T(-2iT_ R^3\Sigma)U_b^c+\textrm{h.c.},
\end{eqnarray}

\noindent where $(Q,Q')$ and $(U,U')$ are multiplets of $SO(6)_A$ and $SO(6)_B$, respectively, defined by:

\begin{eqnarray}
 Q^T&=&\frac{1}{\sqrt{2}}\Big[-(Q_{a1}+Q_{b2}),i(Q_{a1}-Q_{b2}),\\\nonumber
 &&(Q_{a2}-Q_{b1}),i(Q_{a2}+Q_{b1}),Q_{5},Q_6\Big],
\end{eqnarray}
where $(Q_{a1},Q_{a2})$ and $(Q_{b1},Q_{b2})$ represent $SU(2)_L$ doublets, and $(Q_5,Q_6)$ are singlets under $SU(2)_L \times SU(2)_R = SO(4)$. While
\begin{eqnarray}
 (U^c)^T&=&\frac{1}{\sqrt{2}}\Big[-(U^c_{b1}+U^c_{a2}),i(U^c_{b1}-U^c_{a2}),\\\nonumber
 &&(U^c_{b2}-U^c_{a1}),i(U^c_{b2}+U^c_{a1}),U^c_5,U^c_6\Big],
\end{eqnarray}
where $(U^c_{a2},-U^c_{a1})$ and $(-U^c_{b2},U^c_{b1})$ represent doublets of $SU(2)_L$ along with the singlets $(U_5,U_6)$. And
\begin{eqnarray}
 Q^{\prime T}_{a}&=&\frac{1}{\sqrt{2}}\left(-Q'_{a1},iQ'_{a1},Q'_{a2},iQ'_{a2},0,0\right)\\
 U^{\prime c T}_5&=&(0,0,0,0,U^{\prime c}_5,0),
\end{eqnarray}
are a doublet of $SU(2)_A$ and a singlet of $SU(2)_{A,B}$, respectively.
$S=\text{diag}(1,1,1,1,-1,-1)$ represents a symmetry operator, $(y_1,y_2,y_3)$ denote Yukawa couplings, and the term $(q_3,U_b^c)$ in Eq. (\ref{lag-yuk}) encodes information regarding the bottom quark.
The BLHM introduces novel physics into the gauge, fermion, and Higgs sectors, leading to the presence of partner particles for the majority of SM particles. Given that top quark loops contribute significantly to the divergent quantum corrections to the Higgs mass in the SM, the new heavy quarks introduced in the BLHM framework are expected to play a pivotal role in addressing the hierarchy problem.
The heavy quarks include $T$, $T^5$, $T^6$, $T^{2/3}$, $T^{5/3}$, and $B$, each with assigned masses~\cite{schmaltz2010bestest}:
\begin{eqnarray}
\label{masa-T}
m_{T}^2&=&(y_1^2+y_2^2)f^2\\\nonumber
&+&\frac{9v_1^2y_1^2y_2^2y_3^2}{(y_1^2+y_2^2)(y_2^2-y_3^2)},\\
\label{masa-T5}
m_{T^{5}}^2&=&(y_1^2+y_3^2)f^2\\\nonumber
&-&\frac{9v_1^2y_1^2y_2^2y_3^2}{(y_1^2+y_3^2)(y_2^2-y_3^2)},\\
\label{masa-T6}
m_{T^{6}}^2&=&m_{T^{2/3}}^2=m_{T^{5/3}}^2=y_1^2f^2,\\
\label{masa-B}
m_{B}^2&=&y_B^2f^2=(y_1^2+y_2^2)f^2.
\end{eqnarray}

\noindent In the Lagrangian of the quark sector~\cite{schmaltz2010bestest}, the Yukawa couplings are constrained to $0<y_i<1$. Additionally, the masses of the top ($t$) and bottom ($b$) quarks are generated by the Yukawa couplings $y_t$ and $y_b$~\cite{phdthesis}.

\begin{eqnarray}
m^2_t&=&y_t^2v_1^2,\label{masa-top}\\
m^2_b&=&y_b^2v_1^2-\frac{2y^2_b}{3\sin^2\beta}\frac{v^4_1}{f^2}.
\end{eqnarray}

The coupling $y_t$,
\begin{equation}\label{acople-yt}
 y_t^2=\frac{9y_1^2y_2^2y_3^2}{(y_1^2+y_2^2)(y_1^2+y_3^2)},
\end{equation}
is part of the measure of fine-tuning in the BLHM \cite{phdthesis}, defined as
\begin{equation}
\label{ajuste-fino}
\Psi=\frac{27f^2}{8\pi^2v^2\lambda_0\cos^2\beta}\frac{|y_1|^2|y_2|^2|y_3|^2}{|y_2|^2-|y_3|^2}\log\frac{|y_1|^2+|y_2|^2}{|y_1|^2+|y_3|^2}.
\end{equation}

\subsection{Flavor mixing in the BLHM}

In the original development of the BLHM \cite{schmaltz2010bestest}, the authors did not include interactions between heavy quarks $(T, T^5, T^6, T^{2/3}, T^{5/3}, B)$ and the light quarks of the SM $(u, c, d, s)$. This omission prevents the calculation of observables like the one proposed in this article. Therefore, we have implemented the extension to the BLHM introduced in the article \cite{Cisneros-Perez2023}. This extension allows us to obtain interactions and contributions from heavy quarks to the chromodipole moments of light quarks. The best way to do this is adding the terms 
\begin{equation}\label{saborE}
    y_Bfq_1(-2iT_R^2\Sigma)d_B^c,\hspace{0.2cm}y_Bfq_2(-2iT_R^2\Sigma)d_B^c,
\end{equation}

\noindent to the Lagrangian \ref{lag-yuk}. Here, $y_B^2=y_1^2+y_2^2$ is the Yukawa coupling of heavy $B$ quark, $q_1$ and $q_2$ are multiplets of light SM quarks and $d_B^c$ is a new multiplet containing the heavy $B$ quark. This modification give us the interactions between the scalar fields $(H^{\pm},\phi^{\pm},\eta^{\pm})$, the heavy quark $B$ and the light SM quarks $(u, c, d, s)$. The vectorial interactions between
the fields $(W^{\pm},W'^{\pm})$, the heavy quark $B$, and the light SM quarks are allowed by adding to the Lagrangian that describes fermion-gauge interactions~\cite{schmaltz2010bestest,phdthesis}, the terms 
\begin{equation}\label{saborV}
    \sum_{i=1}^2i\bar{\sigma}_{\mu}Q^{\dagger}_3D^{\mu}q_i,\hspace{0.5cm}\sum_{i=1}^4i\bar{\sigma}_{\mu}q_i^{\prime\dagger}D^{\mu}U^c,
\end{equation}

\noindent where $\bar{\sigma}_{\mu}=-\sigma_{\mu}$ are the Pauli matrices, $Q_3^T=(1/\sqrt{2})(0,0,B,iB,0,0)$ and $q_i^{\prime T}=(0,0,0,0,q_i^c,0)$ with $i=1,2$. The covariant derivative $D_{\mu}$ contains information about $(W^{\pm},W^{\prime\pm})$. With these changes, we can introduce two extended matrices of the Cabibbo-Kobayashi-Maskawa (CKM) type \cite{kobayashi1973cp}, $V_{Hu}$ and $V_{Hd}$, such that $V_{CKM}=V_{Hu}^{\dagger}V_{Hd}$, where $V_{CKM}$ is the CKM matrix of the SM.
Regarding the interactions of the light quarks with the neutral heavy bosons $(h^0,H^0,A^0,\phi^0,\eta^0,\sigma,Z,Z^{\prime},\gamma)$ and the quark $B$, these are also included both in Eq. \ref{saborE} and Eq. \ref{saborV}.

\section{The CMDM in the BLHM}
\label{seccion3}

The effective Lagrangian describing the contributions of the vertex $g\bar{q}_iq_i$, where $q_i={u,c,d,s}$, is provided by:

\begin{equation}
    \mathcal{L}_{eff}=-\frac{1}{2}\bar{q}_i\sigma^{\mu\nu}\left(\hat{\mu}_{q_i}+i\hat{d}_{q_i}\gamma^5\right)q_iG^a_{\mu\nu}T^a,
\end{equation}

\noindent where $G^a_{\mu\nu}$ represents the gluon field strength tensor, $T^a$ denotes the generators of $SU(3)$, $\hat{\mu}_{q_i}$ stands for the CMDM, and $\hat{d}_{q_i}$ represents the CEDM, such that

\begin{equation}\label{dip-usual}
    \hat{\mu}_{q_i}=\frac{m_{q_i}}{g_s}\mu_{q_i},\hspace{0.5cm}\hat{d}_{q_i}=\frac{m_{q_i}}{g_s}d_{q_i}.
\end{equation}
The definitions provided by Eq.~(\ref{dip-usual}) are standard relations for the CMDM and the CEDM commonly found in the literature, as $\mathcal{L}_{\text{eff}}$ has dimension 5. Here, $m_{q_i}$ denotes the mass of each light quark, and $g_s=\sqrt{4\pi \alpha_s}$ represents the coupling constant of the group. In our scenario, we solely need to evaluate the chromomagnetic form factor $\mu_{q_i}$ originating from one-loop contributions of the scalar fields $A^0, H^0, H^{\pm}, h^0, \sigma, \phi^0, \phi^{\pm},\eta^0, \eta^{\pm}$, the vector fields $Z^0, Z', W^{\pm}, W'^{\pm}$, and the heavy quarks $T, T^5, T^6, T^{2/3}, T^{5/3}, B$. As for the CEDM, it has been demonstrated in \cite{Aranda:2020tox} to be identically zero within the BLHM framework, thus necessitating no further consideration in this study.

\section{Parameter space of the BLHM}
\label{seccion4}

Through the different publications of the BLHM, two types of parameter spaces have been utilized. On one hand, the initial publications on the model \cite{phdthesis,Martin:2012kqb,Godfrey:2012tf,kalyniak2015constraining} parametrized the Yukawa couplings $(y_1, y_2, y_3)$ in terms of two angles $(\theta_{12}, \theta_{13})$, dividing the space into two parts, leading to heavy quark masses acquiring two hierarchies depending on whether $y_2 < y_3$ or $y_3 < y_2$. On the other hand, the parameter space we employ is the same as proposed in \cite{Aranda:2020tox,Cisneros-Perez2023,Cisneros-Perez:2024onx}, as it can be optimized more easily without dividing it into two parts or inducing hierarchies in the masses of heavy quarks. In this implementation, the Yukawa couplings of the BLHM are maintained in the range $0<y_i<1$, ensuring that the relation \ref{acople-yt} is satisfied under the condition \ref{masa-top} and the value $m_t=172.13$ GeV \cite{CMS:2021jnp}. We solve Eq.(\ref{masa-H0}) to deduce the masses of the scalar bosons in the model considering $0.15\leq\beta\leq1.49$ radians and $m_{h^0}=125.46$ GeV \cite{CMS:2020xrn} under the condition $\lambda^0<4\pi$ \cite{kalyniak2015constraining}. The authors of BLHM impose the condition $\tan\beta>1$, ensuring that the contributions from radiative corrections at one-loop from the top quark and heavy tops to the Higgs mass are minimized. This narrows down the interval for $\beta$ to $0.79<\beta<1.49$ radians. According to the constraints deduced in \cite{Cisneros-Perez:2024onx} to maintain the fine-tuning $\Psi$, Eq. \ref{ajuste-fino}, within the interval $0<\Psi<2$, we also adopt the same values for all parameters of the BLHM as shown in Table \ref{masas-esc1}. From this table divided into minimums and maximums according to the different intervals of the breaking scale $f$, we observe the allowed masses for the scalar fields $A^0$, $H^0$, and $H^{\pm}$.
The parameter $\alpha$ is the mixing angle between the fields $h^0$ and $H^0$ \cite{phdthesis}, such that the alignment limit is satisfied, $\cos(\beta-\alpha)\approx0$, \cite{hashemi2022}.
The mass of the scalar boson $\sigma$  is the largest among the scalars of the BLHM and is given by the expression $m_{\sigma}^2=2\lambda^0K_{\sigma}f^2$,\cite{kalyniak2015constraining}, that is equivalent to Eq. \ref{masa-sigma} but more easy to calculate. For the scalar $\eta^0=m_4$, where $m_4$ is a free parameter of the model \cite{schmaltz2010bestest}, we choose the range $30 \leq m_4 \leq 800$ GeV \cite{Cisneros-Perez2023}  due to the growing magnitudes of masses for experimentally sought new particles. In the case of the charged scalar bosons  $\phi^{\pm}$, $\eta^{\pm}$ and the neutral $\phi^0$, their masses also depend on $m_4$ as well as on both breaking scales $f$ and $F$ and one loop contributions from the Coleman-Weinberg potential \cite{schmaltz2010bestest,Martin:2012kqb}. We can see their values in the range $1\leq f\leq3$ TeV and $F=5$ TeV in Table \ref{masas-boson}.

\begin{table}[H]
\begin{center}
\caption {Scalar masses of the BLHM. } \label{masas-boson}
\medskip
\begin{tabular}{|c|c|c|c|}
\hline
\multirow{2}{*}{Mass\hspace{3mm}} & \multicolumn{2}{c}{Values} & \\
\cline{2-4}
  & $f(1\,{\scriptstyle\text{TeV}})$ & $f(3\,{\scriptstyle\text{TeV}})$  & Unit\\
\hline
\rule{0pt}{3ex}$m_{\sigma}$                          & 1414.2    & 4242.6    & GeV    \\
$m_{\phi^0}$                          & 836.1    & 999.3    & GeV    \\
$m_{\phi^{\pm}}$                         & 841.9   & 1031.9      & GeV  \\
$m_{\eta^{\pm}}$                            & 580.0    & 1013.9      & GeV   \\
\hline
\end{tabular}
\end{center}
\end{table}

\begin{table*}
\caption {Parameters and scalar masses  constrained in the BLHM.} \label{masas-esc1}
\medskip
\begin{tabular}{|c|c|c|c|c|c|c|c|}
\hline
\multirow{2}{*}{Parameter } & \multicolumn{2}{c}{$f(1\,{\scriptstyle\text{TeV}})$}  & \multicolumn{2}{c}{$f(2\,{\scriptstyle\text{TeV}})$} & \multicolumn{2}{c}{$f(3\,{\scriptstyle\text{TeV}})$}&\\
\cline{2-8}
  & \hspace{3mm}Min\hspace{3mm} & \hspace{3mm}Max\hspace{3mm} & \hspace{3mm}Min\hspace{3mm} & \hspace{3mm}Max\hspace{3mm} & \hspace{3mm}Min\hspace{3mm}  & \hspace{3mm}Max\hspace{3mm} & \hspace{3mm}Unit\hspace{3mm}  \\
\hline
\rule{0pt}{3ex}$\beta$                          & 0.79    & 1.47       & 0.79   & 1.36  &  0.79 & 1.24 & rad\\
$\alpha$                         & -0.99    & 0.00       & -0.99   & -0.16  & -0.99 & -0.31 & rad\\
$\Psi$                           & 0.096   & 2.11       & 0.38   & 2.03  & 0.87  & 2.04 & -- \\
$m_{A^0}$                        & 125.0  & 884.86     & 125.0   & 322.75   & 125.0  & 207.07  & GeV\\
$m_{H^0}$                        & 872.04  & 1236.06     & 872.04   & 921.42  & 872.04  & 887.53  & GeV\\
$m_{H^{\pm}}$                    & 125.0  & 884.86     & 125.0   & 322.75   & 125.0  & 207.07  & GeV\\
\hline
\end{tabular}
\end{table*}

The masses of the heavy vector bosons $(W^{\prime\pm},Z^{\prime})$ depend on the scales $f$ and $F=5$ TeV. To determine their masses, we use Eqs. \ref{masa-zp} and \ref{masa-wp}, which are shown in Table \ref{masas-vector}.

\begin{table}[H]
\begin{center}
\caption {Vector masses of the BLHM. } \label{masas-vector}
\medskip
\begin{tabular}{|c|c|c|c|}
\hline
\multirow{2}{*}{Mass\hspace{3mm}} & \multicolumn{2}{c}{Values} & \\
\cline{2-4}
  & $f(1\,{\scriptstyle\text{TeV}})$ & $f(3\,{\scriptstyle\text{TeV}})$  & Unit\\
\hline
\rule{0pt}{3ex}$m_{W^{\prime\pm}}$       & 3328.63    & 3806.44    & GeV    \\
$m_{Z^{\prime}}$              & 3327.65    & 3805.58    & GeV    \\
\hline
\end{tabular}
\end{center}
\end{table}

The masses of the six heavy quarks introduced in this model are given by Eqs.~(\ref{masa-T})-(\ref{masa-B}), where we do not see a dependence on the angle $\beta$ but only on the Yukawa couplings $(y_1,y_2,y_3)$ and the breaking scale $f$. This allows for simpler calculation of these masses in the interval $1<f<3$ TeV; their values are shown in Table \ref{masas-quarks}.

\begin{table}[H]
\begin{center}
\caption {Quarks masses of the BLHM. } \label{masas-quarks}
\medskip
\begin{tabular}{|c|c|c|c|}
\hline
\multirow{2}{*}{Mass\hspace{3mm}} & \multicolumn{2}{c}{\hspace{3mm}$1\leq f\leq 3$ {\scriptsize TeV}} & \\
\cline{2-4}
  & Min & Max  & Unit\\
\hline
\rule{0pt}{3ex}$m_{T}$                          & 1140.18    & 3420.53    & GeV    \\
$m_{T^5}$                          & 773.88    & 2321.66    & GeV    \\
$m_{T^6}$                         & 780.0   & 2100.0      & GeV  \\
$m_{T^{2/3}}$                            & 780.0    & 2100.0      & GeV   \\
$m_{T^{5/3}}$                            & 780.0    & 2100.0      & GeV   \\
$m_{B}$                            & 1140.18    & 3420.53      & GeV   \\
\hline
\end{tabular}
\end{center}
\end{table}

\begin{figure*}
\includegraphics[scale=0.5]{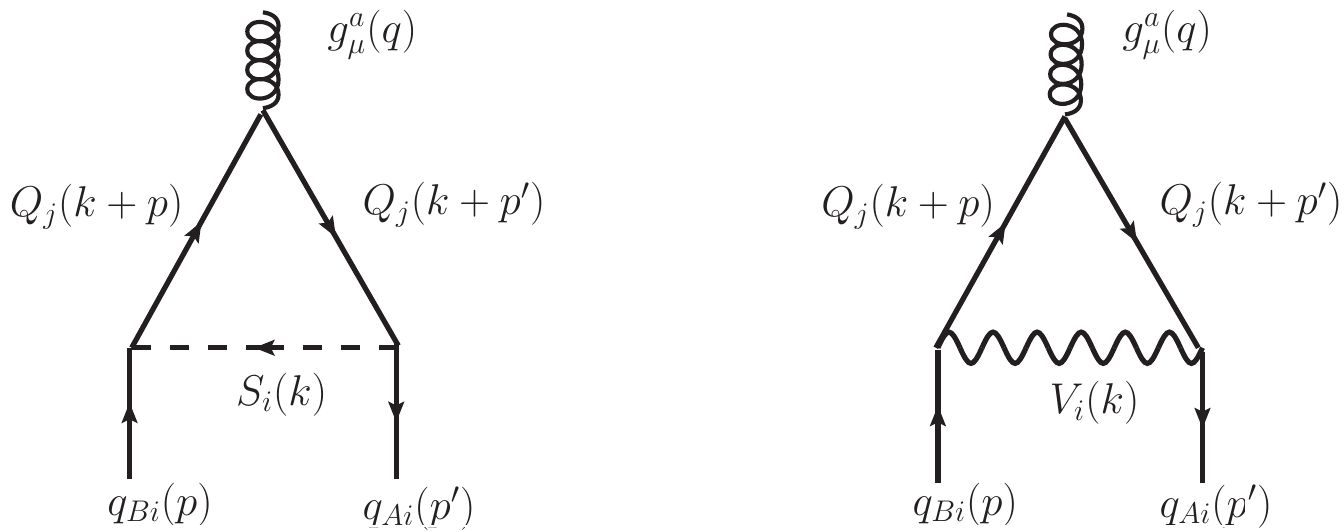}
\caption{In the left diagram, the interactions of the SM light quarks $(u,c,d,s,b)$ and the BLHM heavy quarks $Q=(T,T^5,T^6,T^{2/3},T^{5/3},B)$ with scalar fields $S_i$ are depicted: $A^0, H^0, h^0, H^{\pm}, \phi^0, \eta^0, \sigma, \phi^{\pm}, \eta^{\pm}$. In the right diagram, the interactions of the same quarks with vector fields $V_i$ are illustrated: $Z^0, W^{\pm}, \gamma, Z^{\prime}, W^{\prime\pm}$.
}
 \label{dipolo}
\end{figure*}

\subsection{Experimental limits for the BLHM}

Experimentally, the pursuit of a heavy neutral scalar, such as $A^0$ and $H^0$, is in line with the mass range of the BLHM. In \cite{ATLAS:2020gxx}, masses ranging from $230-800$ GeV for $m_{A^0}$ and $130-700$ GeV for $m_{H^0}$ are explored in the decay $A\to ZH$, based on an integrated luminosity of $139$ fb$^{-1}$ from pp collisions at $\sqrt{s} = 13$ TeV recorded by the ATLAS detector and interpreted within the 2HDM framework. The study by ATLAS \cite{ATLAS:2015kpj} analyzes the process $A\to Zh$, excluding masses of $A^0$ below 1 TeV at 95\% C.L. for all types of 2HDM. Similarly, in \cite{CMS:2019ogx} at CMS, masses of $A^0$ below 1 TeV are also ruled out. In \cite{hashemi2019search}, type I 2HDMs are investigated through the simulation of the process $e^{-} e^{+} \to AH$ using the SiD detector at the ILC, with an integrated luminosity of 500 fb$^{-1}$. This results in ranges of $200 < m_{A^0} < 250$ GeV and $150 < m_{H^0} < 250$ GeV. For $H^{\pm}$, the study in \cite{cms2022search} examines the process $H^{\pm}\to HW^{\pm}$ at CMS in pp collisions at $\sqrt{s} = 13$ TeV with an integrated luminosity of $138$ fb$^{-1}$, considering $m_{H^{\pm}}$ in the range $300-700$ GeV. Additionally, in \cite{ATLAS:2021upq}, the decay $H^{+}\to t\bar{b}$ is explored in pp collisions at $\sqrt{s}=13$ TeV with 139 fb$^{-1}$ of data at ATLAS, with $m_{H^{+}}$ considered in the range $200-2000$ GeV. The exploration for heavy Higgs bosons appears to be highly active across various channels and theoretical frameworks, such as the 2HDM. Furthermore, all the mass ranges either encompass or are encompassed by those investigated in this study.

The presence of neutral and charged fields, denoted by $(\phi, \eta)$, derived from pseudo Goldstone bosons, is a common feature shared with other LHM frameworks and proposals for dark matter. However, experimental searches primarily focus on fields associated with the Higgs rather than scalars of this nature \cite{PhysRevD.107.055022,Buttazzo:2015txu,Haisch:2021ugv}. Another distinctive characteristic of the BLHM is the real scalar field $\sigma$, which is anticipated to be the heaviest among the scalars. Nevertheless, its contribution to the CMDM of light SM quarks is nearly negligible due to certain constraints imposed by the CKM extended matrix.
Within the domain of heavy quarks, the decay channels $T\to Ht$ or $T\to Zt$ are scrutinized in \cite{ATLAS:2023pja}. This investigation examines proton-proton collisions at $\sqrt{s}=13$ TeV with an integrated luminosity of 139 fb$^{-1}$ at ATLAS, revealing no significant signals at the 95\% confidence level for the mass of the $T$ in the range of $1.6-2.3$ TeV. Analogous searches for $T$ and $B$ can be found in \cite{ATLAS:2022tla,ATLAS:2018cye,CMS:2018zkf}. In \cite{ATLAS:2022tla}, they also explore the potential decay of a quark with charge $5/3$, such as $T^{5/3}$, to $Wt$, establishing a lower limit for $m_{T^{5/3}}$ of $1.42$ TeV. For the BLHM within our parameter space, we find $m_{T^{5/3}}=m_{T^6}=m_{T^{2/3}}$ spanning the range $780-2100$ GeV, as shown in Table \ref{masas-quarks}. In the BLHM, we introduce additional vector bosons $W^{\prime\pm}$ and $Z^{\prime}$, whose masses are constrained to be equal based on the properties chosen for our parameter space. Several investigations into the existence of the $W^{\prime}$ boson have been reported. In \cite{CMS:2022ncp}, various mass ranges for $m_{W'}$ are considered within the theoretical framework of different extended models, with values ranging from $2.2$ to $4.8$ TeV. As for the $Z^{\prime}$ boson, recent searches indicate its mass to be above $4.7$ TeV \cite{CMS:2021klu} and within the range of $800-3700$ GeV \cite{CMS:2021fyk}.

\section{Phenomenology of the CMDM of the light quarks}
\label{seccion5}

Regarding the CMDM, the permissible one-loop diagrams involving scalar and vector contributions are depicted in Fig. \ref{dipolo}. The amplitudes associated with the dipole diagrams for each $(u, c, d, s, b)$ quark are outlined as follows:
\begin{eqnarray}\label{ampE}\\\nonumber
  &&\mathcal{M}^{\mu}_{q_n}(S)=\sum_{i,j}\int\frac{d^4k}{(2\pi)^4}\bar{u}_n(p^{\prime})(S^{\ast}_{nj}+P^{\ast}_{nj}\gamma^5)\delta_{A\alpha_1}\\\nonumber
  &&\times\left[i\frac{\slashed{k}+\slashed{p}^{\prime}+m_{Q_j}}{(k+p^{\prime})^2-m_{Q_j}^2}\delta_{\alpha_1\alpha_3}\right]\left(-ig_s\gamma^{\mu}T^a_{\alpha_2\alpha_3}\right)\\\nonumber
  &&\times\left[i\frac{\slashed{k}+\slashed{p}+m_{Q_j}}{(k+p)^2-m^2_{Q_j}}\right](S_{nj}+P_{nj}\gamma^5)\delta_{B\alpha_4}u_n(p)\\\nonumber
  &&\times\left(\frac{i}{k^2-m^2_{S_i}}\right)V_{H_jq_n}^{\ast}V_{H_jq_n}
\end{eqnarray}
\noindent and
\begin{eqnarray}\label{ampV}\\\nonumber
  &&\mathcal{M}^{\mu}_{q_n}(V)=\sum_{i,j}\int\frac{d^4k}{(2\pi)^4}\bar{u}_n(p^{\prime})\gamma^{a_1}(V^{\ast}_{nj}+A^{\ast}_{nj}\gamma^5)\delta_{A\alpha_1}\\\nonumber
  &&\times\left[i\frac{\slashed{k}+\slashed{p}^{\prime}+m_{Q_j}}{(k+p^{\prime})^2-m_{Q_j}^2}\delta_{\alpha_1\alpha_3}\right]\left(-ig_s\gamma^{\mu}T^a_{\alpha_2\alpha_3}\right)\\\nonumber
  &&\times\left[i\frac{\slashed{k}+\slashed{p}+m_{Q_j}}{(k+p)^2-m^2_{Q_j}}\right]\gamma^{a_2}(V_{nj}+A_{nj}\gamma^5)\delta_{B\alpha_4}u_n(p)\\\nonumber
  &&\times\left[\frac{i}{k^2-m^2_{V_i}}\left(-g_{\alpha_1\alpha_2}+\frac{k_{\alpha_1}k_{\alpha_2}}{m^2_{V_i}}\right)\right]V^{\ast}_{H_jq_n}V_{H_jq_n},
\end{eqnarray}

\noindent where $T^a_{\alpha_n\alpha_m}$ represent the generators of $SU(3)$. The coefficients $(S_{nj},P_{nj},V_{nj},A_{nj})$ encompass all contributions from the BLHM, quantified by the vertices $\bar{Q}_jS_iq_n$, $\bar{q}_nS_i^{\dagger}Q_j$ for scalar and pseudoscalar interactions, and $\bar{Q}_jV_iq_n$, $\bar{q}_nV_i^{\dagger}Q_j$ for vector and axial interactions, respectively. The matrix elements $V^{\ast}_{H_jq_n}V_{H_jq_n}$ pertain to the extended CKM matrix. The amplitudes (\ref{ampE}) and (\ref{ampV}) were computed utilizing the \texttt{FeynCalc} package \cite{shtabovenko2020feyncalc} and the \texttt{Package X} \cite{patel2015package} for \texttt{Mathematica}.

In interactions involving charged bosons, both scalar and vector, the extended CKM matrix for the BLHM, denoted as $V_{CKM}=V_{Hu}^{\dagger}V_{Hd}$, as introduced in \cite{Cisneros-Perez2023}, must be taken into account. Here, the unitary matrix $V_{Hu}^{\dagger}$ represents transitions from heavy quarks to light up-type quarks, while $V_{Hd}$ represents transitions from heavy quarks to light down-type quarks. The CKM extended matrix can be conceptualized as the product of three rotation matrices, \cite{blanke2007another,blanke2007rare},

\begin{eqnarray}\label{VHd}
V_{Hd}\;&&=\begin{pmatrix}
1&0&0\\
0&c_{23}^d&s_{23}e^{-i\delta_{23}^d}\\
0&-s_{23}^de^{i\delta_{23}^d} &c_{23}^d
\end{pmatrix}\\\nonumber
&&\times \begin{pmatrix}
c_{13}^d&0&s_{13}^de^{-i\delta_{13}^d}\\
0&1&0\\
-s_{13}e^{i\delta_{13}^d}&0&c_{13}^d
\end{pmatrix}\\\nonumber
&&\times \begin{pmatrix}
c_{12}^d&s_{12}^de^{-i\delta_{12}^d}&0\\
-s_{12}^de^{i\delta_{12}^d}&c_{12}^d&0\\
0&0&1
\end{pmatrix},
\end{eqnarray}

\noindent where the parameters $c^d_{ij}$ and $s^d_{ij}$ are expressed in relation to the angles $(\theta_{12},\theta_{23},\theta_{13})$ and the phases $(\delta_{12},\delta_{23},\delta_{13})$.

In \cite{Cisneros-Perez2023} and \cite{Cisneros-Perez:2024onx}, the authors choose three cases to parameterize the matrices $V_{Hu}$ and $V_{Hd}$ in such a way that results with greater variation could be obtained. However, in \cite{Cisneros-Perez:2024onx}, they find practically the same behavior for all three cases of the extended CKM matrices. For this reason, in this study, we have chosen to use only the third case since we also calculate the CMDM.

We construct the extended CKM matrix as follows:
Substituting the values $s_{23}^d=1/\sqrt{2}$, $s_{12}^d=s_{13}^d=0$, $\delta_{12}^d=\delta_{23}^d=\delta_{13}^d=0$ into the matrix $V_{Hd}$ in Eq.~(\ref{VHd}), we obtain the matrix:

\begin{equation}\label{VHd3}
 V_{Hd}=
 \begin{pmatrix}
  1 & 0 & 0\\
  0 & 1 & 1/\sqrt{2}\\
  0 & -1/\sqrt{2} & 1
 \end{pmatrix},
 \end{equation}
\\
 \noindent and through the product $V_{Hd}V_{CKM}^{\dagger}$, we obtain the matrix
\begin{equation}
\label{vhu3}
V_{Hu}=
\begin{aligned}
& \left(
\begin{matrix}
0.974  & 0.225  & 0.008  \\
0.151  & 0.668  & 0.497  \\
-0.103  & -0.431  & 0.646 
\end{matrix}
\right)
\end{aligned}
\end{equation}

\subsection{Results}

We have computed the CMDMs of the light quarks $(u, c, d, s, b)$ under one-loop contributions from heavy quarks and bosons in the BLHM. We have also included interactions with the fields $(h^0,Z, \gamma, W^{\pm})$ of the SM. Interactions with the charged bosons $(W^{\pm}, W^{\prime\pm},H^{\pm}, \phi^{\pm}, \eta^{\pm})$ are mediated by the matrix elements of $V_{Hu}$ and $V_{Hd}$, Eqs. \ref{vhu3} and \ref{VHd3} respectively. Since the virtual quarks in the dipole could only be the quarks from the BLHM, certain constraints emerged for the model's contributions towards the light quarks of the SM. One of them arose with the charged bosons because in most valid vertices, interaction was only with the heavy quark $B$ (see Tables \ref{feynrulesTabla1} and \ref{feynrulesTabla2} in Appendix \ref{feynrules}). Although the heavy quarks $(T, T^5, T^6, T^{2/3}, T^{5/3})$ do not interact with the light quarks $(u, c, d, s)$, the contributions from the BLHM,  due also to the interactions with the fields $(H^0,A^0,\phi^0,\eta^0,\sigma,Z^{\prime})$, enhance their CMDMs compared to those of the SM. We have evaluated the CMDM of the light quarks taking the gluon off-shell $(q^2\neq0)$ in two scenarios: the spacelike $(q^2=-m^2_Z)$ and the timelike one $(q^2=m^2_Z)$. By solving the amplitudes \ref{ampE} and \ref{ampV}, expressions of the magnetic form factor in terms of the Passarino-Veltman functions of the type $A_0$, $B_0$, and $C_0$ are obtained.

\begin{figure}[H]
 \centering
 \includegraphics[width=8.5cm]{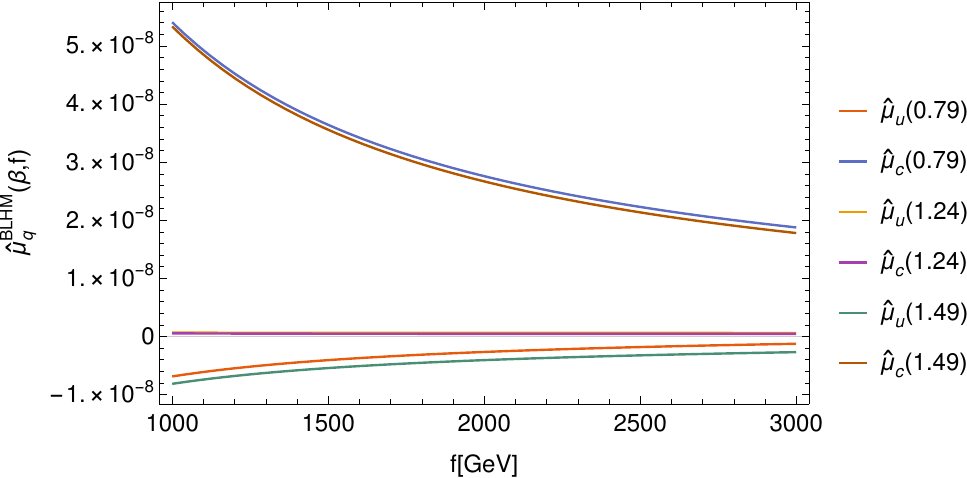}
 \caption{Dipole of the $u$ and $c$ quarks in the spacelike scenario for $\beta=(0.79,1.24,1.49)$ radians, $1<f<3$ TeV and $F=5$ TeV.}
 \label{UP-CHARM-s}
\end{figure}

\begin{figure}[H]
 \centering
 \includegraphics[width=8.5cm]{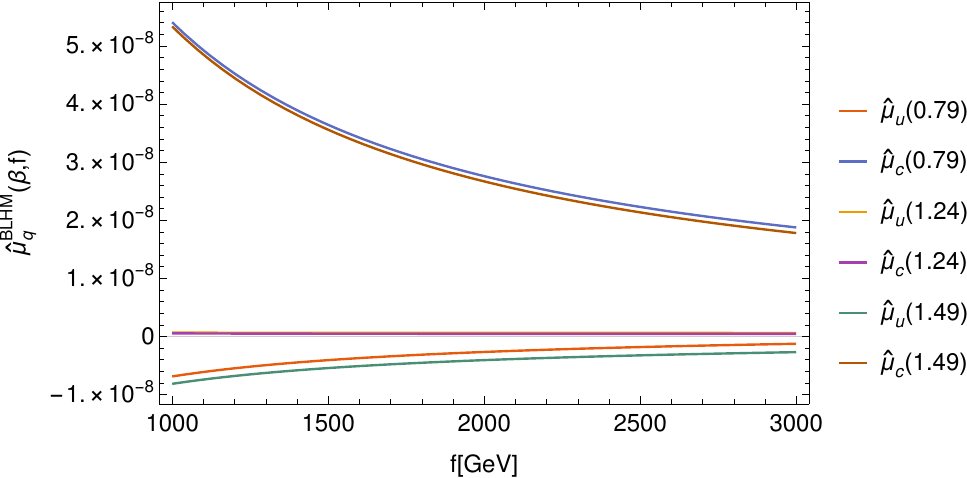}
 \caption{Dipole of the $u$ and $c$ quarks in the timelike scenario with $\beta=(0.79,1.24,1.49)$ radians, $1<f<3$ TeV and $F=5$ TeV.}
 \label{UP-CHARM-t}
\end{figure}

Figs. \ref{UP-CHARM-s} and \ref{UP-CHARM-t} contains the CMDM graphics of the up and charm quarks. Due to the BLHM contributions to the dipoles of these two quarks having similar magnitudes, $\mathcal{O}(10^{-8})$, we have plotted them together for $\beta=(0.79,1.24,1.49)$ radians. It is important to note that they only receive contributions from the $B$ quark, both with charged and neutral bosons.

\begin{figure}[H]
 \centering
 \includegraphics[width=8.5cm]{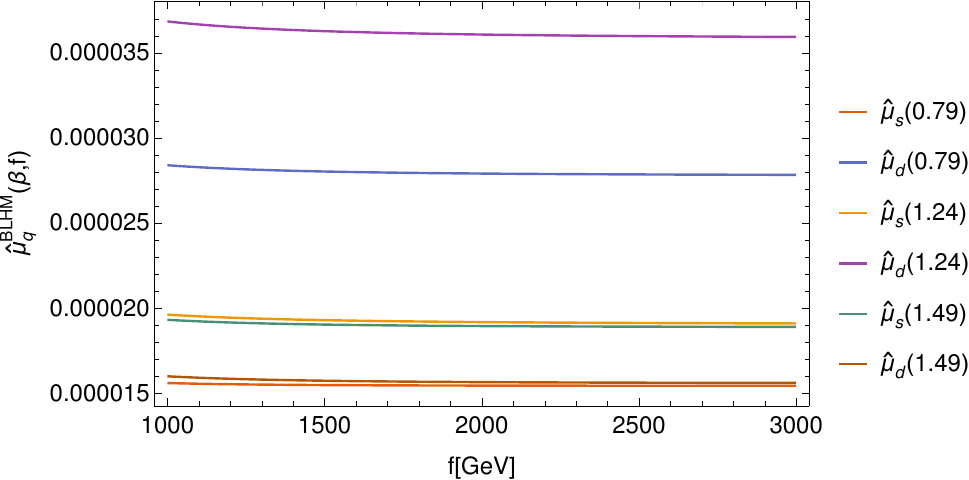}
 \caption{Dipole of the $s$ and $d$ quarks in the spacelike scenario with $\beta=(0.79,1.24,1.49)$ radians, $1<f<3$ TeV and $F=5$ TeV.}
 \label{STRANGE-DOWN-s}
\end{figure}

\begin{figure}[H]
 \centering
 \includegraphics[width=8.5cm]{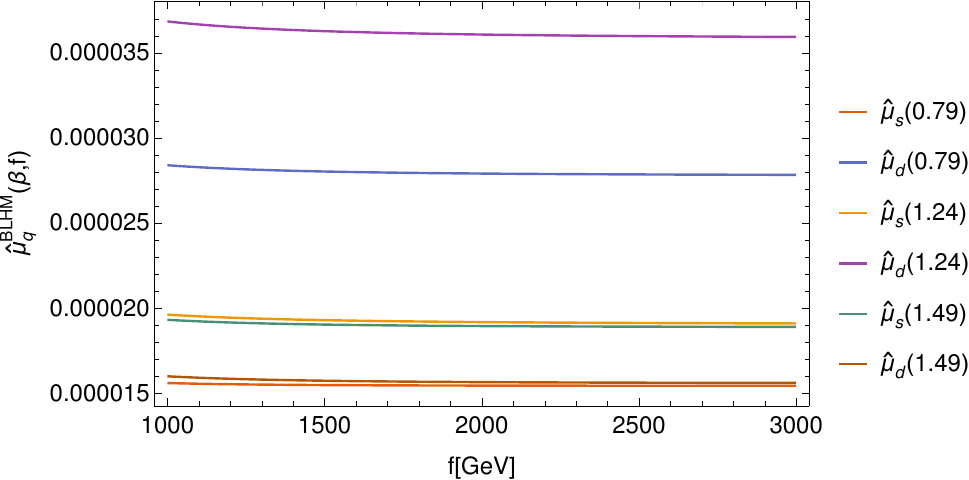}
 \caption{Dipole of the $s$ and $d$ quarks in the timelike scenario with $\beta=(0.79,1.24,1.49)$ radians, $1<f<3$ TeV and $F=5$ TeV.}
 \label{STRANGE-DOWN-t}
\end{figure}

Figs. \ref{STRANGE-DOWN-s} and \ref{STRANGE-DOWN-t} depict the dipoles of the strange and down quarks, respectively, for the same mentioned angles. They receive contributions of the order $10^{-5}$, both in spacelike and timelike configurations.

\begin{figure}[H]
 \centering
 \includegraphics[width=8.5cm]{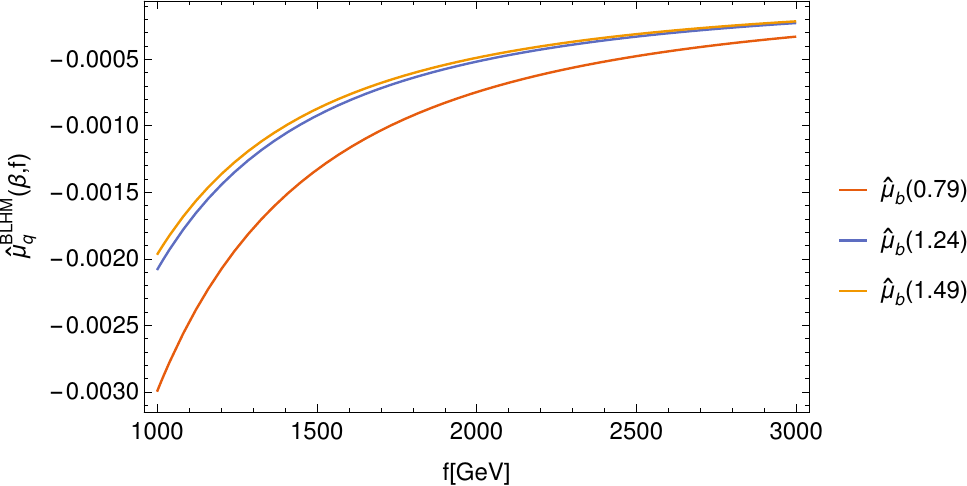}
 \caption{Dipole of the $b$ quark in the spacelike scenario with $\beta=(0.79,1.24,1.49)$ radians, $1<f<3$ TeV and $F=5$ TeV.}
 \label{BOTTOM-s}
\end{figure}

\begin{figure}[h]
 \centering
 \includegraphics[width=8.5cm]{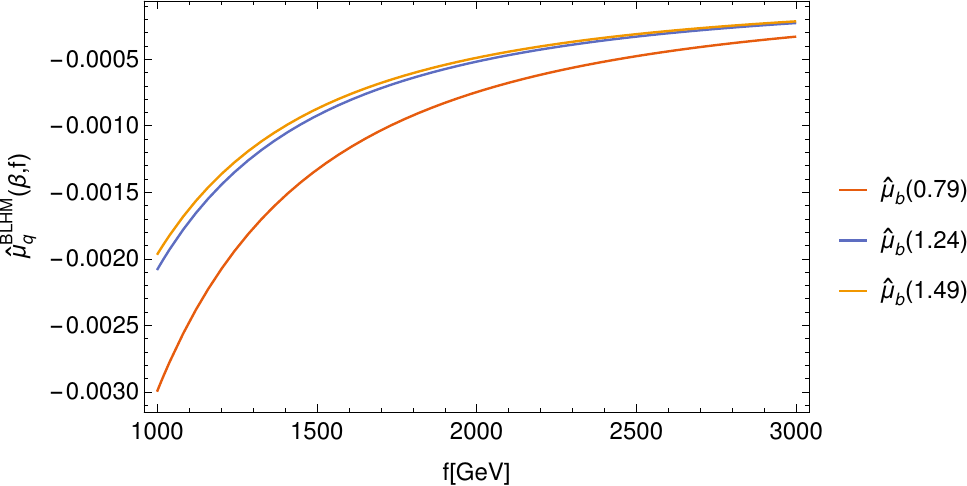}
 \caption{Dipole of the $b$ quark in the timelike scenario with $\beta=(0.79,1.24,1.49)$ radians, $1<f<3$ TeV and $F=5$ TeV.}
 \label{BOTTOM-t}
\end{figure}

Figs. \ref{BOTTOM-s} and \ref{BOTTOM-t} show us the improved contributions from the BLHM quarks and bosons to the SM bottom quark. Both spacelike and timelike CMDMs are of the order of $10^{-3}$. These results indicate that it is necessary to add new interactions between heavy quarks of the $T$ type with the light quarks $(u, c, d, s)$.

\begin{table}[H]
\begin{center}
\caption {Numerical values for $\hat{\mu}_{q_i}^{\scaleto{BLHM}{3pt}}(\beta,f)$ spacelike at $f=1$ TeV and $F=5$ TeV. } \label{dipolos-N1s}
\medskip
\begin{tabular}{|c|c|c|c|}
\hline
 \rule{0pt}{3ex}$\beta$\hspace{2mm} & 0.79 & 1.24  & 1.49\\
\hline
\rule{0pt}{3ex}$\hat{\mu}_u^{\scaleto{BLHM}{3pt}}$\hspace{2mm}    & $-6.89\times10^{-9}$    & $6.45\times10^{-10}$    & $-8.16\times10^{-9}$    \\
\rule{0pt}{3ex}$\hat{\mu}_c^{\scaleto{BLHM}{3pt}}$\hspace{2mm}     & $5.4\times10^{-8}$    & $5.14\times10^{-10}$    & $5.33\times10^{-8}$    \\
\rule{0pt}{3ex}$\hat{\mu}_s^{\scaleto{BLHM}{3pt}}$\hspace{2mm}     & $1.55\times10^{-5}$   & $1.46\times10^{-5}$      & $1.93\times10^{-5}$  \\
\rule{0pt}{3ex}$\hat{\mu}_d^{\scaleto{BLHM}{3pt}}$\hspace{2mm}     & $2.83\times10^{-5}$    & $3.68\times10^{-5}$      & $1.59\times10^{-5}$   \\
\rule{0pt}{3ex}$\hat{\mu}_b^{\scaleto{BLHM}{3pt}}$\hspace{2mm}     & $-2.99\times10^{-3}$    & $-2.08\times10^{-3}$      & $-1.97\times10^{-3}$   \\
\hline
\end{tabular}
\end{center}
\end{table}

\begin{table}[H]
\begin{center}
\caption {Numerical values for $\hat{\mu}_{q_i}^{\scaleto{BLHM}{3pt}}(\beta,f)$ timelike at $f=1$ TeV and $F=5$ TeV. } \label{dipolos-N1t}
\medskip
\begin{tabular}{|c|c|c|c|}
\hline
 \rule{0pt}{3ex}$\beta$\hspace{2mm} & 0.79 & 1.24  & 1.49\\
\hline
\rule{0pt}{3ex}$\hat{\mu}_{u}^{\scaleto{BLHM}{3pt}}$\hspace{2mm}    & $-6.89\times10^{-9}$    & $6.45\times10^{-10}$    & $-8.16\times10^{-9}$    \\
\rule{0pt}{3ex}$\hat{\mu}_{c}^{\scaleto{BLHM}{3pt}}$\hspace{2mm}     & $5.4\times10^{-8}$    & $5.14\times10^{-10}$    & $5.33\times10^{-8}$    \\
\rule{0pt}{3ex}$\hat{\mu}_s^{\scaleto{BLHM}{3pt}}$\hspace{2mm}     & $1.55\times10^{-5}$   & $1.46\times10^{-5}$      & $1.93\times10^{-5}$  \\
\rule{0pt}{3ex}$\hat{\mu}_d^{\scaleto{BLHM}{3pt}}$\hspace{2mm}     & $2.83\times10^{-5}$    & $3.68\times10^{-5}$      & $1.59\times10^{-5}$   \\
\rule{0pt}{3ex}$\hat{\mu}_b^{\scaleto{BLHM}{3pt}}$\hspace{2mm}     & $-2.99\times10^{-3}$    & $-2.08\times10^{-3}$      & $-1.97\times10^{-3}$   \\
\hline
\end{tabular}
\end{center}
\end{table}

\begin{table}[H]
\begin{center}
\caption {Numerical values for $\hat{\mu}_{q_i}^{\scaleto{BLHM}{3pt}}(\beta,f)$ spacelike at $f=2$ TeV and $F=5$ TeV. } \label{dipolos-N2s}
\medskip
\begin{tabular}{|c|c|c|c|}
\hline
 \rule{0pt}{3ex}$\beta$\hspace{2mm} & 0.79 & 1.24  & 1.49\\
\hline
\rule{0pt}{3ex}$\hat{\mu}_u^{\scaleto{BLHM}{3pt}}$\hspace{2mm}    & $-2.64\times10^{-9}$    & $-3.58\times10^{-9}$    & $-4.12\times10^{-9}$    \\
\rule{0pt}{3ex}$\hat{\mu}_c^{\scaleto{BLHM}{3pt}}$\hspace{2mm}     & $-5.15\times10^{-8}$    & $-5.22\times10^{-8}$    & $-5.26\times10^{-8}$    \\
\rule{0pt}{3ex}$\hat{\mu}_s^{\scaleto{BLHM}{3pt}}$\hspace{2mm}     & $1.53\times10^{-5}$   & $1.89\times10^{-5}$      & $1.88\times10^{-5}$  \\
\rule{0pt}{3ex}$\hat{\mu}_d^{\scaleto{BLHM}{3pt}}$\hspace{2mm}     & $2.56\times10^{-5}$    & $3.11\times10^{-5}$      & $1.67\times10^{-5}$   \\
\rule{0pt}{3ex}$\hat{\mu}_b^{\scaleto{BLHM}{3pt}}$\hspace{2mm}     & $-6.0\times10^{-4}$    & $-4.37\times10^{-4}$      & $-5.0\times10^{-4}$   \\
\hline
\end{tabular}
\end{center}
\end{table}

\begin{table}[H]
\begin{center}
\caption {Numerical values for $\hat{\mu}_{q_i}^{\scaleto{BLHM}{3pt}}(\beta,f)$ timelike at $f=2$ TeV and $F=5$ TeV. } \label{dipolos-N2t}
\medskip
\begin{tabular}{|c|c|c|c|}
\hline
 \rule{0pt}{3ex}$\beta$\hspace{2mm} & 0.79 & 1.24  & 1.49\\
\hline
\rule{0pt}{3ex}$\hat{\mu}_u^{\scaleto{BLHM}{3pt}}$\hspace{2mm}    & $-2.68\times10^{-9}$    & $5.95\times10^{-10}$    & $-4.08\times10^{-9}$    \\
\rule{0pt}{3ex}$\hat{\mu}_c^{\scaleto{BLHM}{3pt}}$\hspace{2mm}     & $2.75\times10^{-8}$    & $4.34\times10^{-10}$    & $2.66\times10^{-8}$    \\
\rule{0pt}{3ex}$\hat{\mu}_s^{\scaleto{BLHM}{3pt}}$\hspace{2mm}     & $1.54\times10^{-5}$   & $1.91\times10^{-5}$      & $1.89\times10^{-5}$  \\
\rule{0pt}{3ex}$\hat{\mu}_d^{\scaleto{BLHM}{3pt}}$\hspace{2mm}     & $2.78\times10^{-5}$    & $3.60\times10^{-5}$      & $1.56\times10^{-5}$   \\
\rule{0pt}{3ex}$\hat{\mu}_b^{\scaleto{BLHM}{3pt}}$\hspace{2mm}     & $-7.48\times10^{-4}$    & $-5.19\times10^{-4}$      & $-4.9\times10^{-4}$   \\
\hline
\end{tabular}
\end{center}
\end{table}

\begin{table}[H]
\begin{center}
\caption {Numerical values for $\hat{\mu}_{q_i}^{\scaleto{BLHM}{3pt}}(\beta,f)$ spacelike at $f=3$ TeV and $F=5$ TeV. } \label{dipolos-N3s}
\medskip
\begin{tabular}{|c|c|c|c|}
\hline
 \rule{0pt}{3ex}$\beta$\hspace{2mm} & 0.79 & 1.24  & 1.49\\
\hline
\rule{0pt}{3ex}$\hat{\mu}_u^{\scaleto{BLHM}{3pt}}$\hspace{2mm}    & $-1.25\times10^{-9}$    & $-2.20\times10^{-9}$    & $-2.74\times10^{-9}$    \\
\rule{0pt}{3ex}$\hat{\mu}_c^{\scaleto{BLHM}{3pt}}$\hspace{2mm}     & $-3.39\times10^{-8}$    & $-3.46\times10^{-8}$    & $-3.50\times10^{-8}$    \\
\rule{0pt}{3ex}$\hat{\mu}_s^{\scaleto{BLHM}{3pt}}$\hspace{2mm}     & $1.53\times10^{-5}$   & $1.88\times10^{-5}$      & $1.87\times10^{-5}$  \\
\rule{0pt}{3ex}$\hat{\mu}_d^{\scaleto{BLHM}{3pt}}$\hspace{2mm}     & $2.56\times10^{-5}$    & $3.10\times10^{-5}$      & $1.66\times10^{-5}$   \\
\rule{0pt}{3ex}$\hat{\mu}_b^{\scaleto{BLHM}{3pt}}$\hspace{2mm}     & $-2.69\times10^{-4}$    & $-1.93\times10^{-4}$      & $-2.20\times10^{-4}$   \\
\hline
\end{tabular}
\end{center}
\end{table}

\begin{table}[H]
\begin{center}
\caption {Numerical values for $\hat{\mu}_{q_i}^{\scaleto{BLHM}{3pt}}(\beta,f)$ timelike at $f=3$ TeV and $F=5$ TeV. } \label{dipolos-N3t}
\medskip
\begin{tabular}{|c|c|c|c|}
\hline
 \rule{0pt}{3ex}$\beta$\hspace{2mm} & 0.79 & 1.24  & 1.49\\
\hline
\rule{0pt}{3ex}$\hat{\mu}_u^{\scaleto{BLHM}{3pt}}$\hspace{2mm}    & $-1.27\times10^{-9}$    & $5.86\times10^{-10}$    & $-2.71\times10^{-9}$    \\
\rule{0pt}{3ex}$\hat{\mu}_c^{\scaleto{BLHM}{3pt}}$\hspace{2mm}     & $1.87\times10^{-8}$    & $4.19\times10^{-10}$    & $1.77\times10^{-8}$    \\
\rule{0pt}{3ex}$\hat{\mu}_s^{\scaleto{BLHM}{3pt}}$\hspace{2mm}     & $1.54\times10^{-5}$   & $1.90\times10^{-5}$      & $1.88\times10^{-5}$  \\
\rule{0pt}{3ex}$\hat{\mu}_d^{\scaleto{BLHM}{3pt}}$\hspace{2mm}     & $2.78\times10^{-5}$    & $3.59\times10^{-5}$      & $1.55\times10^{-5}$   \\
\rule{0pt}{3ex}$\hat{\mu}_b^{\scaleto{BLHM}{3pt}}$\hspace{2mm}     & $-3.31\times10^{-4}$    & $-2.29\times10^{-4}$      & $-2.16\times10^{-4}$   \\
\hline
\end{tabular}
\end{center}
\end{table}

In the plots from Figs. \ref{UP-CHARM-s} to \ref{BOTTOM-t}, no differences are observed between the dipoles evaluated in the spacelike and timelike scenarios. Except for the dipoles when $f=1$ TeV, as shown in Tables \ref{dipolos-N1s} and \ref{dipolos-N1t}, the dipoles for $f=2$ TeV and $f=3$ TeV do exhibit differences, which we present in Tables \ref{dipolos-N2s} to \ref{dipolos-N3t}. Also, we have included the CMDMs of the light quarks calculated in the SM (see Table \ref{dipolos-SM}). As we can observe, if we compare them with Tables \ref{dipolos-N1s} and \ref{dipolos-N1t}, only $\hat{\mu}_c^{\scaleto{BLHM}{3pt}}$ is lower than its counterpart in the SM. On the other hand, $\hat{\mu}_b^{\scaleto{BLHM}{3pt}}$ exceeds the order of $\hat{\mu}_b^{SM}$. According to the charm mass, one might expect the contributions to the dipole to be of a magnitude close to that of the bottom, but they turned out to be very small. This shows us that the contributions of the quarks $(T,T^5,T^6,T^{2/3})$ are much more significant than those of all the heavy bosons since the order of $\hat{\mu}_b^{\scaleto{BLHM}{3pt}}$ is much higher due to the heavy up-type quarks.

\begin{table}[H]
\begin{center}
\caption {Numerical values for $\hat{\mu}_{q_i}^{SM}$ in the SM, spacelike and timelike \cite{Montano-Dominguez:2021eeg}. } \label{dipolos-SM}
\medskip
\begin{tabular}{|c|c|c|}
\hline
 \rule{0pt}{3ex} & $q^2=-m_Z^2$ & $q^2=m_Z^2$  \\
\hline
\rule{0pt}{3ex}$\hat{\mu}_u^{SM}$\hspace{2mm}    & $-1.15\times10^{-10}$    & $-1.15\times10^{-10}$        \\
\rule{0pt}{3ex}$\hat{\mu}_c^{SM}$\hspace{2mm}     & $-1.16\times10^{-5}$    & $1.15\times10^{-5}$         \\
\rule{0pt}{3ex}$\hat{\mu}_s^{SM}$\hspace{2mm}     & $-1.38\times10^{-7}$   & $1.37\times10^{-7}$        \\
\rule{0pt}{3ex}$\hat{\mu}_d^{SM}$\hspace{2mm}     & $-5.07\times10^{-10}$    & $-5.04\times10^{-10}$       \\
\rule{0pt}{3ex}$\hat{\mu}_b^{SM}$\hspace{2mm}     & $-1.61\times10^{-4}$    & $1.55\times10^{-4}$         \\
\hline
\end{tabular}
\end{center}
\end{table}

\section{Conclusions}
\label{seccion6}

We have calculated the dipoles of the light quarks of the SM $(u, c, d, s, b)$ with contributions from heavy quarks and bosons of the BLHM, as well as the bosons $(h^0, Z, \gamma, W^{\pm})$ of the SM, both in the spacelike scenario $(q^2=-m^2_Z)$ and the timelike scenario $(q^2=m^2_Z)$. We find that the magnitude of our results is greater than similar ones in the SM. The dipole for which we have the largest contributions from the BLHM is the CMDM of the bottom quark, $\mathcal{O}(10^{-3})$, which receives contributions from all heavy quarks and bosons. On the other hand, the dipoles of the other light quarks $(u, c, d, s)$ only receive contributions from the heavy quark $B$ and heavy bosons \cite{Cisneros-Perez2023} as can be observe from Feynman rules in Tables \ref{feynrulesTabla1} and \ref{feynrulesTabla2}. From this, we deduce that it is also necessary to extend interactions of the heavy up-type quarks to all four light quarks $(u, c, d, s)$. In this study, we have also used the extended CKM matrices $V_{Hu}$ and $V_{Hd}$, parametrizing only one case of them in relation to the $V_{CKM}$ matrix. The elements of the extended matrices have further constrained the dipoles $\hat{\mu}_{q_i}$ when compared to the dipoles calculated for the top quark in \cite{Cisneros-Perez:2024onx}, also in the BLHM. The magnitudes of the CMDMs calculated in the BLHM, mainly for the bottom quark, encourage us to expect new experimental signals that may provide hints of new physics. It is important to remember that the CMDMs or CEDMs of light quarks have not been evaluated in other BSM models as we did in this article, but we hope that they will be done to enrich our own work on this topic.

\begin{acknowledgments}
T. C. P. and E. C. A. thanks SNII and CONAHCYT (México) postdoctoral fellowships.
\end{acknowledgments}

\appendix

\section{Feynman rules in the BLHM}
\label{feynrules}

In this appendix, we derive and present the Feynman rules for the BLHM necessary to calculate the CMDM of light quarks. The Feynman rules for the $b$ quark and other fields can be found in references \cite{Aranda:2021kza, Cruz-Albaro:2023pah, Cisneros-Perez2023}.

Tables \ref{feynrulesTabla1} and \ref{feynrulesTabla2} summarize the Feynman rules for the 3-point interactions: fermion-fermion-scalar (FFS) and fermion-fermion-gauge (FFV) interactions.

\begin{table}[H]
\centering
\caption{Essential Feynman rules in the BLHM for studying CMDM of light quarks are the 3-point interactions fermion-fermion-scalar (FFS) and fermion-fermion-gauge (FFV) interactions.}
\begin{tabular}{|c|c|}\cline{1-2}
\rule{0pt}{4ex} Vertex & Rule \\\cline{1-2}
\rule{0pt}{4ex} $W^{\prime -}\bar{B}u$ & $\dfrac{igg_A}{2\sqrt{2}g_B}\gamma^{\mu}P_L(V_{Hu})$ \\\cline{1-2}
\rule{0pt}{4ex} $W^{\prime -}\bar{B}c$ & $\dfrac{igg_A}{2\sqrt{2}g_B}\gamma^{\mu}P_L(V_{Hu})$ \\\cline{1-2}
\rule{0pt}{4ex} $\eta^{-}\bar{B}u$ & $-\dfrac{im_Bs_{\beta}^2}{2f\sqrt{2}}P_L(V_{Hu})$ \\\cline{1-2}
\rule{0pt}{4ex} $\eta^{-}\bar{B}c$ & $-\dfrac{m_Bs_{\beta}^2}{2f\sqrt{2}}P_L(V_{Hu})$ \\\cline{1-2}
\rule{0pt}{4ex} $\eta^{0}\bar{B}d$ & $-\dfrac{im_Bs_{\beta}^2}{4f}P_L$ \\\cline{1-2}
\rule{0pt}{4ex} $\eta^{0}\bar{B}s$ & $-\dfrac{im_Bs_{\beta}^2}{4f}P_L$ \\\cline{1-2}
\rule{0pt}{4ex} $\phi^{-}\bar{B}u$ & $\dfrac{iFs^2_{\beta}\left[m_u+m_B+(m_u-m_B)\gamma^5\right]}{2f\sqrt{2}\sqrt{f^2+F^2}}(V_{Hu})$ \\\cline{1-2}
\rule{0pt}{4ex} $\phi^{-}\bar{B}c$ & $\dfrac{iFs^2_{\beta}\left[m_c+m_B+(m_c-m_B)\gamma^5\right]}{2f\sqrt{2}\sqrt{f^2+F^2}}(V_{Hu})$ \\\cline{1-2}
\rule{0pt}{4ex} $\phi^{0}\bar{B}d$ & $\dfrac{iFm_Bs^2_{\beta}}{4f\sqrt{f^2+F^2}}P_L$ \\\cline{1-2}
\rule{0pt}{4ex} $\phi^{0}\bar{B}s$ & $\dfrac{iFm_Bs^2_{\beta}}{4f\sqrt{f^2+F^2}}P_L$ \\\cline{1-2}
\rule{0pt}{4ex} $H^{-}\bar{B}u$ & $\dfrac{gm_{B}s_{2\beta}}{4\sqrt{2}m_W}P_L(V_{Hu})$ \\\cline{1-2}
\rule{0pt}{4ex} $H^{-}\bar{B}c$ & $\dfrac{gm_{B}s_{2\beta}}{4\sqrt{2}m_W}P_L(V_{Hu})$ \\\cline{1-2}
\rule{0pt}{4ex} $H^{0}\bar{B}d$ & $-\dfrac{gm_{B}s_{\alpha}s_{\beta}}{4m_W}P_L$ \\\cline{1-2}
\rule{0pt}{4ex} $H^{0}\bar{B}s$ & $-\dfrac{gm_{B}s_{\alpha}s_{\beta}}{4m_W}P_L$ \\\cline{1-2}
\end{tabular}
\label{feynrulesTabla1}
\end{table}

\begin{table}[H]
\centering
\caption{Essential Feynman rules in the BLHM for studying CMDM of light quarks are the 3-point interactions fermion-fermion-scalar (FFS) and fermion-fermion-gauge (FFV) interactions.}
\begin{tabular}{|c|c|}\cline{1-2}
\rule{0pt}{4ex} Vertex & Rule \\\cline{1-2}
\rule{0pt}{4ex} $h^{0}\bar{B}d$ &\hspace{1cm} $\dfrac{gm_{B}c_{\alpha}s_{\beta}}{4m_W}P_L$ \\\cline{1-2}
\rule{0pt}{4ex} $h^{0}\bar{B}s$ &\hspace{1cm} $\dfrac{gm_{B}c_{\alpha}s_{\beta}}{4m_W}P_L$ \\\cline{1-2}
\rule{0pt}{4ex} $A^{0}\bar{B}d$ &\hspace{1cm} $-\dfrac{gm_{B}s_{2\beta}}{8m_W}P_L$ \\\cline{1-2}
\rule{0pt}{4ex} $A^{0}\bar{B}s$ &\hspace{1cm} $-\dfrac{gm_{B}s_{2\beta}}{8m_W}P_L$ \\\cline{1-2}
\rule{0pt}{4ex} $\sigma\bar{B}d$ &\hspace{1cm} $\dfrac{m_{B}s_{2\beta}}{4\sqrt{2}m_Wf}P_L$ \\\cline{1-2}
\rule{0pt}{4ex} $\sigma\bar{B}s$ &\hspace{1cm} $\dfrac{m_{B}s_{2\beta}}{4\sqrt{2}m_Wf}P_L$ \\\cline{1-2}
\rule{0pt}{4ex} $Z\bar{B}d$ &\hspace{1cm} $-\dfrac{ig^2s_W}{4g^{\prime}}\gamma^{\mu}P_L$ \\\cline{1-2}
\rule{0pt}{4ex} $Z\bar{B}s$ &\hspace{1cm} $-\dfrac{ig^2s_W}{4g^{\prime}}\gamma^{\mu}P_L$ \\\cline{1-2}
\rule{0pt}{4ex} $Z^{\prime}\bar{B}d$ &\hspace{1cm} $-\dfrac{igg_A}{4g_B}\gamma^{\mu}P_L$ \\\cline{1-2}
\rule{0pt}{4ex} $Z^{\prime}\bar{B}s$ &\hspace{1cm} $-\dfrac{igg_A}{4g_B}\gamma^{\mu}P_L$ \\\cline{1-2}
\rule{0pt}{4ex} $\gamma\bar{B}d$ &\hspace{1cm} $-\dfrac{igs_W}{4}\gamma^{\mu}$ \\\cline{1-2}
\rule{0pt}{4ex} $\gamma\bar{B}s$ &\hspace{1cm} $-\dfrac{igs_W}{4}\gamma^{\mu}$ \\\cline{1-2}
\end{tabular}
\label{feynrulesTabla2}
\end{table}

\end{document}